\documentclass[12pt,onecolumn,draftclsnofoot]{IEEEtran}
\usepackage{cite} % Written by Donald Arseneau
\usepackage{graphicx} % Written by David Carlisle and Sebastian Rahtz

\usepackage{subfigure} % Written by Steven Douglas Cochran
\usepackage{url} % Written by Donald Arseneau
\usepackage{amssymb}
\usepackage{amsfonts}
\usepackage{amsmath} % From the American Mathematical Society
\begin{document}
\title{Decoding the Golden Code: a VLSI design}
\author{Barbara~Cerato, {\em Student Member, IEEE},~ Guido~Masera, {\em Member,
IEEE}\\ and Emanuele~Viterbo, {\em Senior Member, IEEE}}
\maketitle

\begin{abstract}
Multiple-input multiple-output (MIMO) systems are among the most
promising transmission techniques to achieve high data rate and high
reliability transmission over wireless channels. The recently
proposed Golden code is an optimal space-time block code for $2\times 2$
MIMO systems. The aim of this work is the design of a VLSI decoder for
a MIMO system coded with the Golden code. The architecture is based
on a rearrangement of the sphere decoding algorithm that achieves
maximum-likelihood (ML) decoding performance. Compared to other approaces,
the proposed  solution exhibits an inherent flexibility in terms of
%supported modulation schemes
QAM modulation size and this makes our architecture 
particularly suitable for adaptive modulation schemes.
Relying on the flexibility of this
approach two different architectures are proposed: a
parametric one able to achieve high decoding throughputs
($>$165 Mbps) while keeping low overall decoder complexity (45 KGates),
 with respect to other proposed solutions; a flexible
implementation able to dynamically adapt to the modulation scheme
(4-,16-,64-QAM) retaining the low complexity and high throughput
features. In addition, a deep analysis of finite precision effects
on the performance is presented in this work for both 16 and 64 QAM.

\end{abstract}


% Will be ignored for other modes.
\IEEEpeerreviewmaketitle

\section{Introduction}
% \PARstart{A}{demo} file is ....
% \PARstart{A}{}demo file is ....
% \PARstart{T}{his demo} file is ....
\PARstart{T}{he} hardware implementation of high data rate and high
reliability wireless communication systems is one of the most
widely investigated topics within the scientific community and has
raised new engineering and research challenges for many years.
Higher transmission reliability demands for higher levels of
processing complexity in the mobile terminal, while faster data
rates require increased throughput: both evolutive trends are
strong driving forces for the search of novel efficient
architectures implementing the most critical base-band processing
functions. In particular new standards proposed to regulate Wireless
Local Area Networks (WLAN) and Metropolitan Area Networks (MAN) are
significant examples of very challenging applications from the
implementation point of view.

%{D}{esigning} very high data rate wireless channels with improved quality of
%service is one of nowadays greatest engineering and research challenges.
%Interest in wireless communications is proved by the increasing number of
%standards to regulate Wireless Local Area Networks (LAN) and Metropolitan
%Area Networks (MAN) (802.11 and 802.16 are just two of many examples).

There are two main objectives on which research is actually
focused. The first goal is to make wireless communication data rate comparable
to that
of wired communications: recent results show it is possible to approach 1Gb/s
data rate\cite{gigabit, gigabit2} . The second one is to improve reliability,
by combating multipath,
noise and interference effects. The recourse to multiple-input multiple-
output (MIMO) systems seems to be one of the most promising solutions to reach both
these results.

Traditionally, MIMO systems were conceived with the purpose of dealing with one of these two
objectives, by means of transmit antenna diversity combined with space-time
coding.
More recently great efforts have been made in unifying both goals and
some new space-time codes are now able to reach the best tradeoff between data rate
and diversity gain, although they require more sophisticated detection schemes
at the receiver \cite{LD-STBC, CODES2, CODES3, Golden1, Golden2, Golden3}.

The main contribution of this work is in the hardware design of a
decoder for this kind of codes, in particular for the decoding of a
$2\times 2$ MIMO signal coded with the {\em Golden code}
\cite{Golden1}. Golden code is a recently proposed full-rate and
full-diversity space-time block code, chosen for its good energy
efficiency. The maximum-likelihood (ML) decoding algorithm for the Golden
code is based on the \emph{Sphere Decoder}, which has already been
widely addressed in the literature also from a hardware
implementation point of view~\cite{ControCompl}-\cite{Kbest}.

Several architectures have been proposed for the efficient
implementation of the sphere decoding architecture, but they are
optimized for specific modulation schemes and do not support
reconfigurability features. In \cite[ASIC-I]{ethvlsi}, in order
to reach high throughput dedicated multipliers and parallel
computations are used adopting a so called ``one node per cycle''
architecture. Other architectures %\cite{lundvlsi} 
instead take advantage of suboptimal algorithms: good examples of this
approach are given in \cite[ASIC-II]{ethvlsi},
where the $L_\infty$-norm is implemented as an alternative to the more expensive
$L_2$-norm, and in~\cite{lundvlsi}, where the K-Best algorithm allows for
performance-complexity trade-offs.
These choices lead to fully optimized architectures,
achieving high throughput; however, they are not ML 
(the loss is about 1.4 dB in the case of  \cite[ASIC-II]{ethvlsi}) and have
been proposed for specific modulation and transmission schemes, 
although in~\cite{ethvlsi} the possibility to adapt the proposed solution to 
different modulations is also mentioned.
In this work we
overcome these limitations, proposing two novel architectures
designed with VHDL as a reusable intellectual property (IP)
macrocell: the first one is parametrized with respect to the
fixed-point representation of data and to the addressed modulation
scheme; in order to enable comparisons with previous
implementations, synthesis results are provided for this
architecture in the case of 16 QAM. The second architecture is
flexible, meaning that it can be dynamically configured to cover
multiple modulation schemes. We note that both these hardware
implementations can be equivalently used in a $4\times 4$ uncoded MIMO system~\cite{gigabit2}.

 In Section \ref{GC} we briefly explain properties,
construction and detection of Golden code, Section \ref{SD} is
dedicated to reviewing the sphere decoding algorithm. In
section \ref{FIX} the effects of fixed-point precision on the code
performance are derived. A short introduction to the overall scheme
of the MIMO receiver is given in Section \ref{Preproc}, with
particular attention to QR decomposition preprocessing unit; the
detailed descriptions of the two hardware implementations are then
carried out in \ref{HW1} and \ref{HW2}. In the last two sections
results and conclusions are presented.


%\hfill November 18, 2002

%%%%%%%%%%%%%%%%%%%%%%%%%%%%%%%%%%%%%%%%%%%%%%%%%%%%%%%%%%%%%%%%%%%%%%%%%%%%%%%%%%%%%%%%%%
\section{Golden code}
\label{GC}
The Golden code is a space-time (ST) code for a $2\times2$ coherent MIMO channel,
it was found independently by \cite{Golden1, Golden2, Golden3}.

Number theoretical methods have been widely employed to construct
full-rate and full-diversity codes for coherent MIMO systems. These
methods are based on the rank and the determinant criteria. In a
Rayleigh fading channel the pairwise error probability (PWEP)
expression \cite{Criterion} shows that the error probability can be
minimized operating mainly on two aspects: {\em diversity} and {\em
coding gain}. In \cite{Criterion} it was proved that these
parameters are related to the so called codeword difference matrix
${\bf D}$, which is constructed as the difference between two
codewords. In order to maximize the diversity gain, the space-time
code must be designed so that the difference matrix between any two
codewords is full rank (\emph {rank criterion}). On the other hand,
the coding gain, depends on the determinant of ${\bf D}{\bf D}^\dag$
and high coding gain is achieved maximizing the minimum of this
determinant over all codeword pairs (\emph{determinant
criterion}).

Golden code satisfies both the rank and the determinant criterion
and in particular, differently from previously known codes, presents
the non-vanishing determinant property, i.e., its minimum
determinant is 1/5 and does not depend on the size of the signal
constellation. For this reason it can be successfully employed in
systems with adaptive selection of the modulation.

Besides these properties, the Golden code has also the peculiarity
to be energy efficient. It is constructed using a rotated version of
the $\mathbb{Z}[i]^4$  complex lattice, so that there is no loss due
to shaping~\cite{Golden1}.

The codewords ${\bf X}$ of the Golden code are $2\times 2$ complex
matrices of the following form 

\begin{equation}
{\bf X} = \frac{1}{\sqrt 5} \left[ \begin{array}{cc}
     \alpha [a + b\theta ] & \alpha [ c + d\theta ]\\
     i\sigma(\alpha)[c + d\sigma(\theta)] & \sigma(\alpha)[a + b\sigma(\theta)]\\
\end{array} \right]
\end{equation}

\noindent where $a, b, c, d$ are the information symbols chosen in a $Q^2$-QAM=$(Q$-PAM$)^2$ constellation,
$i=\sqrt{-1}, \theta=(1+\sqrt{5})/2 = 1.618 \ldots$ (Golden number), $\sigma(\theta)=
(1-\sqrt{5})/2 = 1-\theta, \alpha=1 + i - i \theta= 1 + i \sigma(\theta),
 \sigma(\alpha) = 1 + i - i \sigma(\theta) = 1 + i\theta$, \cite{GoldenPage}.

\subsection{The $2\times 2$ MIMO System Model}
%% Golden code is applied to 2-transmit 2-receive antenna MIMO systems and requires
%% two channel uses to transmit a codeword. 
In order to model the $2 \times 2$
MIMO channel, its impulse response can be used. Assuming
$h_{ij}$ as the time-varying channel fading coefficients between the
$j$-th transmit antenna and the $i$-th receive antenna, the
MIMO channel is described through a 2$\times$2 matrix:

\begin{equation}
\boldsymbol{\mathcal{H}} = \left[ \begin{array}{ll}
       h_{11} & h_{12} \\
       h_{21} & h_{22} \\
\end{array} \right]
\end{equation}

\noindent where $h_{ij} \sim \mathcal{N}_c(0,1)$.
Assuming the ``Block Fading'' channel model, each transmitted codeword will be
affected by an independently varying channel matrix $\boldsymbol{\mathcal{H}}$.
Then, the $2 \times 2$ received matrix is
\[
{\bf Y} = \boldsymbol{\mathcal{H}} {\bf X} + {\bf Z}
\]
where ${\bf Z}$ is the additive white gaussian noise matrix with entries
$\sim \mathcal{N}_c(0,N_0)$.

We note that each codeword is sent in two channel uses of the two
transmit antennas, for a total of four component signals. It is
convenient to represent the codeword ${\bf X}$ in vectorized form
where, furthermore, real and imaginary components are separated,
resulting in a $8 \times 1$ real vector $\boldsymbol{x}$. The channel
matrix $\boldsymbol{\mathcal H}$ can be consequently rearranged in a
8$\times$8 real-valued matrix $\boldsymbol{H}$. It can be seen that
$\boldsymbol{x} = \boldsymbol{G}\boldsymbol{s}$, where
$\boldsymbol{G}$ is a $8 \times 8$ orthogonal matrix
($\boldsymbol{G}^{-1}=\boldsymbol{G}^T$) and $\boldsymbol{s}=(\Re
a,\Im a,\Re b,\Im b,\Re c,\Im c,\Re d,\Im d)$ with entries from a
$Q$-PAM constellation, \cite{GoldenPage}.

The vectorized system model can so be expressed as:
\begin{equation}
\label{model_Golden}
  \boldsymbol{y}= \boldsymbol{H x}+\boldsymbol{z} = \boldsymbol{HGs}+\boldsymbol{z}
\end{equation}
\noindent where $\boldsymbol{y}$ is the 8$\times$1 received real vector and
$\boldsymbol{z}$ is a 8-dimensional i.i.d. (independent and identically 
distributed) zero
mean gaussian noise real vector.

%%%%%%%%%%%%%%%%%%%%%%%%%%%%%%%%%%%%%%%%%%%%%%%%%%%%%%%%%%%%%%%%%%%%%%%%%%%%%%%%%%%%%%%%%%
\subsection{Decoding the Golden code}
Decoding the Golden code is equivalent to decoding an 8-dimensional lattice
with generator matrix $\boldsymbol{M}=\boldsymbol{HG}$.
Provided that $\boldsymbol{H}$ is perfectly known at the
receiver, the optimal detector for a MIMO channel,
which minimizes the codeword error rate, is the maximum-likelihood (ML) detector.
It solves the following equation:
\begin{eqnarray}
\label{MLeq}
\boldsymbol{\hat{s}} %& = &
& = & \arg \min_{\boldsymbol {s} \in Q^{n}} \| \boldsymbol {y - Ms} \|^2 %\\
\end{eqnarray}

\noindent where $Q^n$ is the cardinality of the search space and $n=8$.

The above expression represents a discrete least-square (LS) minimization problem. Exhaustive search
of the ML solution has exponential complexity and in this particular case it has
$2^{n \log_2 Q}$ possible solutions.
Sphere decoding algorithms have then been proposed in order to decrease
the decoder complexity.

%%%%%%%%%%%%%%%%%%%%%%%%%%%%%%%%%%%%%%%%%%%%%%%%%%%%%%%%%%%%%%%%%%%%%%%%%%%%%%%%%%%%%%%%%%%%%%
\section{Sphere Decoding Algorithm}

\label{SD}

\emph{Sphere decoding algorithms}  denote a family of algorithms,
which aim at lowering the complexity of the
minimization~(\ref{MLeq}) by analyzing only a subset of the solution
space \cite{Caire&ot1}. These algorithms, in a certain range of
parameters which is not too far from those of real systems, show a
polynomial average complexity.
Although other work \cite{ControCompl} denies this theoretical
proof, computer simulations still confirm the practical result. This
behavior is due to the fact that $\boldsymbol{y}$ is not an
arbitrary vector, but it is given by the transmitted vector
$\boldsymbol{Hx}$ with a small offset due to additive
noise $\boldsymbol{z}$.

Sphere decoding algorithms look at the set of  possible solutions as
points of a lattice and try to find the closest point to the
received vector. In particular, a hypersphere is constructed around
the received vector and only points inside it are taken into
account, since the others are actually too far. This constraint can
be written as:
\begin{equation}
\label{sphere_const}
 \| \boldsymbol {y - Ms} \|^2 \le C_0
\end{equation}
where $C_0$ is the square radius of the
hypersphere~\cite{Pohst,Viterbo93,Viterbo99}.  In the following we
describe a method to  easily compute distances between received
signals and lattice points.

\subsubsection{Tree construction}
With a linear transformation of the matrix $\boldsymbol{M}$, such as
QR or Cholesky decomposition, it is possible to rewrite
$\boldsymbol{M}$ as a product of two matrices, one of which upper
triangular \cite{Caire&ot1}. In this work, QR decomposition has been
employed so that, given $\boldsymbol{M}=\boldsymbol{QR}$,
(\ref{MLeq}) can be rewritten as:
\begin{eqnarray}
\label{rw_sphere_const}
\arg \min_{\boldsymbol{s} \in Q^{n}} \left \| \boldsymbol{y - QRs} \right \|^2 & =
& \arg \min_{{\bf s} \in Q^{n}} \left \| \boldsymbol{Q^T y} - \boldsymbol{Rs} \right \|^2 \nonumber\\
& = & \arg \min_{{\bf s} \in Q^{n}} \left \| \boldsymbol{\tilde{y} - Rs} \right
\|^2
\end{eqnarray}
where we have exploited the orthogonality of $\boldsymbol{Q}$ and
$\boldsymbol{\tilde{y}}=\boldsymbol{Q}^T {\bf y}$ represents the
zero-forcing (ZF) solution. The upper triangular structure of the
factored matrix enables to take every component separately into
account for the computation of the distance between the two points.
The distance $d^2(\boldsymbol{s})= \left \| \boldsymbol{\tilde{y} -
Rs} \right \|^2$ can also be computed recursively as follows. Let
us define the partial metric as in~\cite{ethvlsi}
\begin{equation}
\label{MetrEq}
T^{(l)} (\boldsymbol{s}^{(l)}) = \left \{ \begin{array}{l}
0 \qquad \qquad \qquad \qquad \qquad \ \textrm{if $l = n$ + 1}\\
\\
T^{(l+1)}(\boldsymbol {s}^{(l+1)})+ | \tilde{y_l} -
\sum_{j=l}^{n} R_{lj}s_{j}|^2 \\
= T^{(l+1)}(\boldsymbol {s}^{(l+1)})+ | \tilde{y_l} -
\sum_{j=l+1}^{n} R_{lj}s_{j} - R_{ll}s_{l}|^2\\
= T^{(l+1)}(\boldsymbol {s}^{(l+1)})+ | \psi_l^{(l+1)}- R_{ll}s_{l}|^2\\
 \qquad \qquad \qquad \qquad \qquad \quad \textrm{if~}
l=1, ~ \ldots ,~ n
\end{array} \right.
\end{equation}
where $\boldsymbol {s}^{(l)}=[s_l \quad s_{l+1} \quad \dots \quad
s_{n}]$, and
\begin{equation}
\label{Psi} \psi_l^{(l+1)}= \tilde{y_l} - \sum_{j=l+1}^{n}
R_{lj}s_{j}
\end{equation}
\noindent with $\psi_n^{(n+1)}= \tilde{y}_n$. Then we can
write $T^{(1)}(\boldsymbol{s})=d^2(\boldsymbol{s})$.

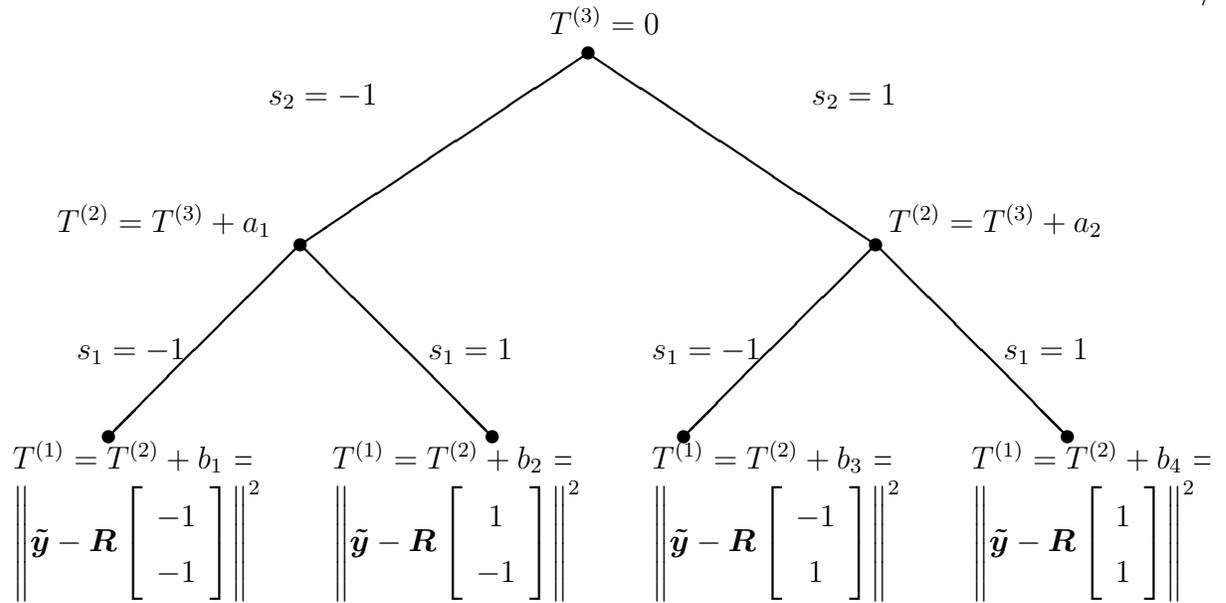
\begin{figure*}[t!]
%\label{Tree}
\begin{center}
\setlength{\unitlength}{0.85mm}
\begin{picture}(150,80) \thicklines%
\put(75,80){\line(-3,-2){45}} %\put(75,60){\vector(-1,-2){15}}
\put(75,80){\line(3,-2){45}} %\put(75,60){\vector(1,-2){15}}
\put(75,80){\circle*{2}}
\put(69,83){$T^{(3)} = 0$}
\put(30,50){\line(-1,-1){30}}
\put(30,50){\line(1,-1){30}}
\put(30,50){\circle*{2}}
\put(120,50){\line(-1,-1){30}}
\put(120,50){\line(1,-1){30}}
\put(120,50){\circle*{2}}
\put(0,20){\circle*{2}}
\put(90,20){\circle*{2}}
\put(60,20){\circle*{2}}
\put(150,20){\circle*{2}}

\put(25,72){$s_2 = -1$}
\put(110,72){$s_2 = 1$}
%\put(15,65){$b(s_2) = |\tilde y_2 + R_{22}|^2$}
%\put(100,65){$b(s_2) = |\tilde y_2 - R_{22}|^2$}

\put(-8,52){$T^{(2)} = T^{(3)} + a_1$}
\put(122,52){$T^{(2)} = T^{(3)} + a_2$}

\put(-5,32){$s_1 = -1$}
\put(50,32){$s_1 = 1$}

\put(85,32){$s_1 = -1$}
%\put(70,38){$b(s_1) = |\psi^2_1 + R_{11}|^2$}

\put(140,32){$s_1 = 1$}

\put(-15,15){$T^{(1)} = T^{(2)} + b_1$ =} \put(-15,2){$\left
\|{\boldsymbol{\tilde y} -\boldsymbol{R} \left [\begin{array}{c}
-1\\
-1\\
\end{array} \right ]} \right \|^2 $}

\put(35,15){$T^{(1)} = T^{(2)} + b_2$ =} \put(35,2){$\left
\|{\boldsymbol{\tilde y} -\boldsymbol{R} \left [\begin{array}{c}
1\\
-1\\
\end{array} \right ]} \right \|^2 $}

\put(85,15){$T^{(1)} = T^{(2)} + b_3$ =} \put(85,2){$\left
\|{\boldsymbol{\tilde y} -\boldsymbol{R} \left [\begin{array}{c}
-1\\
1\\
\end{array} \right ]} \right \|^2 $}

\put(135,15){$T^{(1)} = T^{(2)} + b_4$ =} \put(135,2){$\left
\|{\boldsymbol{\tilde y} -\boldsymbol{R} \left [\begin{array}{c}
1\\
1\\
\end{array} \right ]} \right \|^2 $}

\end{picture}
\end{center}
\caption{Two level tree for QPSK modulation, where $a_1 = |\tilde y_2 + R_{22}|^2$, $a_2 = |\tilde y_2 - R_{22}|^2$, $b_1 = |\tilde y_1 + R_{12} + R_{11}|^2$, $b_2 = ||\tilde y_1 + R_{12} - R_{11} |^2$, $b_3 = |\tilde y_1 - R_{12} + R_{11} |^2$, $b_4 = |\tilde y_1 - R_{12} - R_{11} |^2$}
\label{Tree}
\end{figure*}

One of the most interesting consequences of this interpretation is
that the exploration of the lattice can be thought as a tree
traversal. This tree has $n$ levels and every node at each level has
\emph{Q} sons. At every level the radius
constraint~(\ref{sphere_const}) must be verified and satisfied,
otherwise the branch is pruned. Figure \ref{Tree} depicts a two
level tree for a QPSK modulation. $T^{(l)}$ is the partial distance metric at
level $l$ in (\ref{MetrEq}); at the lowest level, final metrics are explicitly
calculated for this simple case.

\subsubsection{Tree exploration}
Several algorithms have been studied in order to make the tree
traversal efficient. First algorithm, proposed by Pohst in
\cite{Pohst}, needs to chose explicitly an initial radius. This is a
very critical choice: if the radius is chosen too large, too many
points fall into the hypersphere, while for a too small radius no
points are left inside it. A more efficient algorithm has been proposed by
Schnorr and Euchner (SE) \cite{SE}. 
The SE algorithm has intrinsically variable throughput and this makes it
not very suitable for hardware implementation. The key to make this
algorithm efficient or, at least, with predictable throughput, is to
make an effective pruning. A lot of theoretical studies can be found
in recent literature, which aim at finding %a technique 
techniques to reach this
goal \cite{Kbest}. Although some of them give very interesting
ideas, none of them seems to be effective nowadays, with a strong
theoretical demonstration and a simple realization.

%%%%%%%%%%%%%%%%%%%%%%%%%%%%%%%%%%%%%%%%%%%%%%%%%%%%%%%%%%%%%%%%%%%%%%%%%%%%%%%
\section{Fixed-point analysis}
\label{FIX}
The study of finite precision effects is a mandatory preliminary step in the
design and hardware implementation of complex processing tasks.
Although several implementations of the sphere decoding algorithm have been
proposed, studies on finite precision effects are not available in literature.
In this work, a wide range of simulations have been carried out in order to determine the
effects of different fixed-point representations
on the performance for both 16 and 64-QAM modulation schemes.

\begin{figure}[ht]
    \begin{center}
    \includegraphics[angle=270, width=\linewidth]{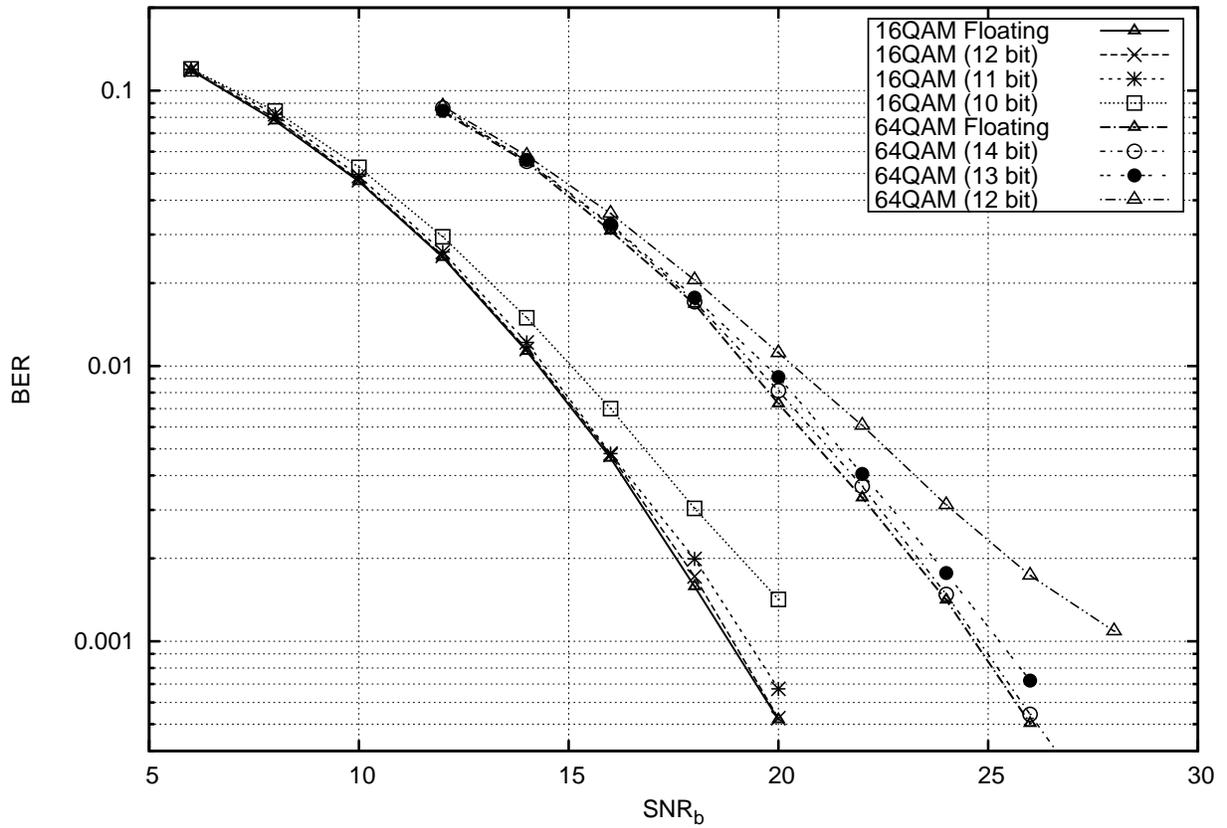}
    \caption{System bit error rate (BER) using 16 and 64-QAM: lowering the total number of bits.}
    \label{BER_Meno}
    \end{center}
\end{figure}

%% A 16 bit data representation has been chosen as a starting point,
%% because it was adopted in most of the literature for the 16-QAM
%% modulation, so it represents a good choice for comparisons. The
%% first objective of these simulations was to find the optimal
%% partitioning between fractional and integer parts within the 16
%% bits.
%% \begin{figure}[ht!]
%%     \begin{center}
%%     \includegraphics[angle=270, width=\linewidth]{ber_Fixed16bitVsFloating.eps}
%%     \caption{System bit error rate (BER) using 16 and 64 QAM: partitioning between integer and fractional part.}
%%     \label{BER_16bit}
%%     \end{center}
%% \end{figure}
%% Figure \ref{BER_16bit} shows the results
%% of our investigation for 16 and 64 QAM cases.
The main conclusion that can be derived from 
results reported in Figure \ref{BER_Meno} is that the
required number of bits increases when
higher--order modulations are used. There are two reasons for
this increase:
\begin{itemize}
\item with higher order modulations, Euclidean distances between
constellation points decrease and a larger number of bits must be 
allocated in the fractional part to discriminate distances; 
\item signal amplitudes are higher in higher order modulations, 
thus more bits need to be allocated also in the integer part.
\end{itemize}
Simulation results show that a total of 12 bits lead to performance 
very close to the floating-point case for 16-QAM modulation,
while 14 bits are necessary in the detection of 64-QAM signals.
%% The best position of the decimal point also depends
%% on the gain exhibited by the whole receiving chain, including RF front-end,
%% analog-to-digital converter and automatic gain control (AGC).

%% More relevant is the total number of bits required to achieve satisfactory BER performance.
%% Further simulations have then been carried out
%% for lower numbers of bits, in order to explore possible trade-offs between hardware cost
%% and detection performance.

\begin{figure}[ht]
    \begin{center}
    \includegraphics[angle=270, width=\linewidth]{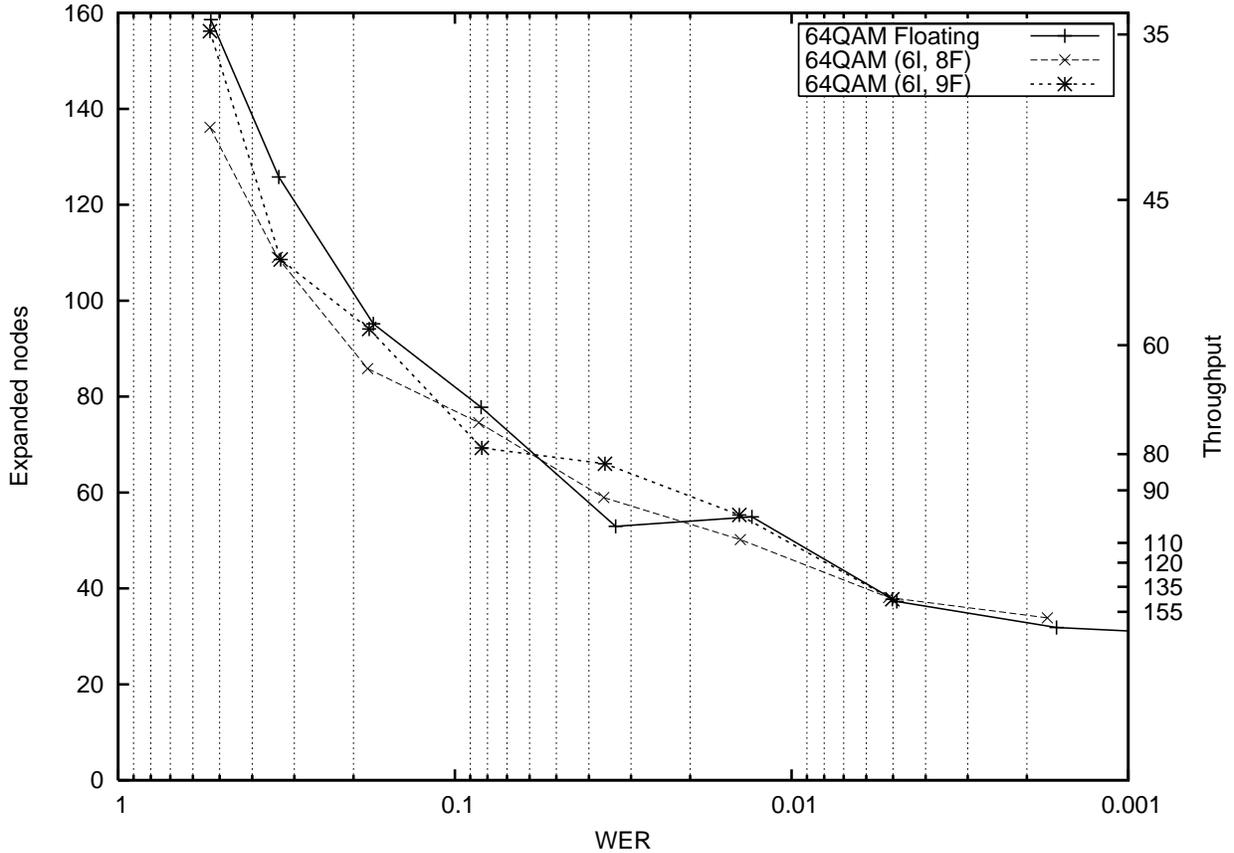}
    \caption{Number of visited nodes using 64 QAM, with different partitioning between integer and fractional part.}
    \label{IT_64QAM_16}
    \end{center}
\end{figure}

Finally, Figures \ref{IT_64QAM_16} %and \ref{IT_64QAM_16} 
shows that the
fixed-point approximation does not affect significantly the number
of visited nodes of the algorithm. The plotis given as a function of the codeword error rate.

%%%%%%%%%%%%%%%%%%%%%%%%%%%%%%%%%%%%%%%%%%%%%%%%%%%%%%%%%%%%%%%%%%%%%%%%%%%%%%%%%
\section{Preprocessing}

\label{Preproc} In this section we discuss the implementation issues
related to pre-processing, which is required before the tree-search.
This computation operates on the lattice generator matrix $\boldsymbol
{M=HG}$; since the code generator matrix is constant, the computation
must be repeated at the channel estimation update frequency.

The update frequency for the channel estimation can change
significantly according to the scenario, but it is generally one or
two orders of magnitude lower than the signalling rate.
\begin{figure*}[t!]
    \begin{center}
    \includegraphics[height=3cm, width=\linewidth ]{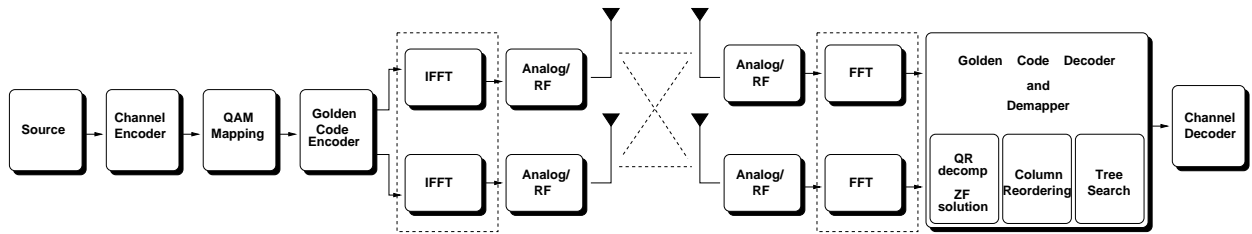} %height=2 cm,
    \caption{Golden Code MIMO System.}
    \label{System}
    \end{center}
\end{figure*}
Figure \ref{System} depicts a block diagram of a MIMO system adopting
the Golden code; dashed blocks implement modulation and
demodulation functions in a generic MIMO-OFDM system. The Golden
code decoding phase is made of three functions: {\em
QR decomposition, column reordering} and {\em tree search}.

While column reordering is an optional operation able to reduce the
tree-search complexity, QR decomposition is mandatory because it
allows constructing the tree and finding the ZF solution, possible
techniques to perform the QR decomposition in hardware are reviewed
in order to estimate the overall complexity of the receiver.

\subsection{QR decomposition}

As already outlined, a linear transformation of the channel matrix $\boldsymbol{H}$,
such as QR or Cholesky decomposition is needed in order to construct the tree.

QR decomposition is a well studied numerical algorithm and widely
used in many applications such as matrix inversion, adaptive
beamforming and filtering. The QR decomposition based - Recursive Least
Squares (QRD-RLS) methods are routinely adopted in applications such
as multiuser detection in CDMA communications, adaptive equalization
of radio channels etc. The method is well suited to VLSI realization
and it can be implemented in a stable manner using relatively short
word length arithmetic.

Hardware realization of this technique implies the choice between
Householder transformation and Givens rotation based algorithms
\cite{MxCom}. This second approach can be accomplished by a sequence
of rotation operations to annihilate elements under the main
diagonal of the matrix. Givens rotations require a larger number of
flops compared to Householder method in order to compute QR
decomposition, nevertheless they may be implemented using highly
parallel systolic arrays and for this reason they are usually
preferred for hardware implementation.

These arrays typically present linear, triangular, or square
structure; the rotation angle is computed in boundary or diagonal
processors and dispatched to other processors for rotation. The
choice of the organization can be made on the basis of area and
throughput considerations. The main parameters of this architecture
are listed in Table \ref{Table: QRcomp}, for a $n \times n$ matrix:
number of {\it processing elements} (PE), latency and throughput. It is
assumed that every processing element takes one or more clock cycles
to perform its computation.

\begin{table*}
\centering
\caption{QR decomposition: different array organization parameters - number of processing elements (PE),
latency and throughput}
\begin{tabular}{|c||c|c|c|} \hline
{\bf Architecture} &{\bf \# of PEs}& {\bf latency of single QR}& {\bf Throughput}\\ \hline \hline
Triangular & $n(n+1)/2$ & $n(n+1)/2$ & $1/n$\\ \hline
 & & $2n^2-n$ & $1/(2n^2-n)$\\
Linear & $n$ & $\div$& $\div$\\
& & $(2n-1)+\left (\frac{n}{2} -1 \right )(n+1)$ & $1/[(2n-1)+\left (\frac{n}{2} -1 \right )(n+1)]$\\ \hline
Single Element & 1 & $n^2(n+1)/2$ & $1/[n^2(n+1)/2]$ \\ \hline\end{tabular}
\label{Table: QRcomp}
\end{table*}

Every single processing element must perform the angle calculation
and the rotation to annihilate the matrix elements. Several
alternatives exist to accomplish these two tasks, and the two main
ones are:

\begin{enumerate}
\item {\it Sine} and {\it cosine} of the angle are computed by means of
operations including also square root and division.
\item Direct calculation of the angle and then rotation using a CORDIC
processor~\cite{CORDIC1}.
\end{enumerate}

The main advantage of the first approach is that primitives can be
optimized resulting in an efficient although expensive
implementation. The second technique is less expensive, but outputs
are generated with longer latencies and data-dependency between
operations slows down the CORDIC algorithm. Many strategies have
been adopted in order to alleviate the effects of data-dependencies, such as
reordering look-ahead\cite{Annih1, Annih2, ReorgCORDIC1} or
redundant arithmetic \cite{RedundantQR}.

For lower data-rates, architectures that reuse the processing elements
on different data have been proposed in~\cite{Alternat1, Alternat2}.
These architectures represent probably the best
tradeoff for the applications addressed in this work.

\section{First Hardware Implementation: \\
Parametrizable Solution}

\label{HW1}

The tree-search algorithm is considered as the most computationally intensive 
processing block in a MIMO detector, although 
column reordering and QR decomposition can also be heavy processing tasks.
However,  since the rate of updating for channel estimation is usually one
or two orders of magnitude lower than the signalling rate, design constraints
tend to be more stringent for the tree-search unit than for
column reordering and the QR decomposition.
Thus the focus of this work is on the hardware realization of
the tree-search algorithm.

As guidelines for the design of the architecture, two main
objectives have been taken into account. The first requirement was a
certain degree of flexibility in the choice of both modulation
scheme. The second main design
objective was a high decoding throughput, compliant with 
needs of modern wireless communication standards.

In the developed architecture, the datapath width, the size
of the search tree and the modulation scheme are tunable
parameters that can be statically configured to make the detector
adaptable to different systems. Although the system is
described with reference to the special case of the Golden code, it
can be also used to decode a $4\times 4$ uncoded MIMO scheme.
The key elements of the developed architecture
are described in the following paragraphs.
\subsection{A flexible hardware solution}

%%%%%%%%%%%%%%%
%% cambio
%%%%%%%%%%%%%%%
The key processing task in the tree exploration algorithm is given by
(\ref{MetrEq}), where we recall that
$\psi_l^{(l+1)}= \tilde{y_l} - \sum_{j=l+1}^{n} R_{lj}s_{j}$,
is the $l$-th entry of an $n$ elements vector 
$\boldsymbol{\psi}^{(l+1)}$, where $l+1$ is the tree level we are referring to. 
At level $l$, the generic $i$-th entry of this vector can be decomposed in a 
recursive manner 
%% \begin{equation}                                                       %%
%% \boldsymbol{\psi}^{(l)}= \boldsymbol{\psi}^{(l+1)}- \boldsymbol{R}_l s_l   %%
%% \label{boldpsieq}                                                         %%
%% \end{equation}                                                            %%
%% \noindent The initial value is given by                                   %%
%% $\boldsymbol{\psi}^{(n+1)}=\boldsymbol{\tilde{y}}$                         %%
%% In particular $\psi_l^{(l)}$ represents the $l$-th entry at level $l$ and the generic $i$-th %%aggiunta
%% entry 
%%can be evaluated %%aggiunta
 through the following expression
\begin{equation}
\label{psiEq}
\psi_i^{(l)} = \left \{ \begin{array}{l}
\tilde{y_i} \qquad \qquad \qquad \qquad \qquad \quad \textrm{if $l = n+1$}\\
\\
\psi_i^{(l+1)} - R_{il} s_{l} \qquad \qquad \qquad  \textrm{if $l = n, \dots , 1$} 
\end{array} \right.
\end{equation}
\noindent where $i$ is in the range $1, \ldots, n$ while the level $l$ decreases from $n+1$ to $1$. 

The whole $\boldsymbol{\psi}^{(l)}$ can therefore be updated by means of 
\begin{equation}                                                       %%
\boldsymbol{\psi}^{(l)}= \boldsymbol{\psi}^{(l+1)}- \boldsymbol{R}_l s_l  ~~~~~~~~~~~~~~~~l = n, \dots , 1 %%
\label{boldpsieq}                                                         %%
\end{equation}   
 \noindent where $\boldsymbol{R}_l$ is the $l$-th column of $\boldsymbol{R}$
and the initial value is given by                                   %%
 $\boldsymbol{\psi}^{(n+1)}=\boldsymbol{\tilde{y}}$.

In order to minimize the
final metric $d^2(\boldsymbol{s})$ with a greedy algorithm, at each level of the tree the
minimum $\psi_l^{(l+1)}-R_{ll}s_l$ value between all sons must be
selected. More precisely,
at each tree node, placed at level $l$, three main operations have to be accomplished:
\begin{enumerate}
\item the $s_l$ that minimizes the difference $|\psi_l^{(l+1)} - R_{ll} s_{l}|$ is selected
\item the partial metric $T^{l}(\boldsymbol{s}^{(l)})$ is calculated according to~(\ref{MetrEq}).
\item for each $i = 1, \ldots, n$, $\psi_i^{(l)}$ is evaluated for the selected $s_l$ value,
according to~(\ref{psiEq})
\end{enumerate}
Thus the straightforward minimization of partial metrics $T^{l}(\boldsymbol{s}^{(l)})$
requires the difference computation for all
the possible values of $s_l$. This technique becomes increasingly expensive 
with high order modulations, due to the large number of required
operations. 

In the proposed architecture, the minimization of $T^{l}(\boldsymbol{s}^{(l)})$
is rearranged in two steps. 
In the first processing step, the value of $s_l$ that minimizes
the difference $|\psi_l^{(l+1)}-R_{ll}s_l|$ is directly selected by means of a division;
the obtained $s_l$ is then used to generate $\psi_i^{(l)}$ amounts in~(\ref{psiEq}),
for all $i = 1, \ldots, n$.
At the second step, 
(\ref{MetrEq}) is finally evaluated  to obtain the actual metric
value $T^{(l)}$ for the selected son.
Two functional blocks, \textbf{U\_psi}
and \textbf{Metric\_compute} units, are allocated to perform the indicated 
processing steps.\\

\begin{figure}[ht]
    \begin{center}
    \includegraphics[height=8.0cm]{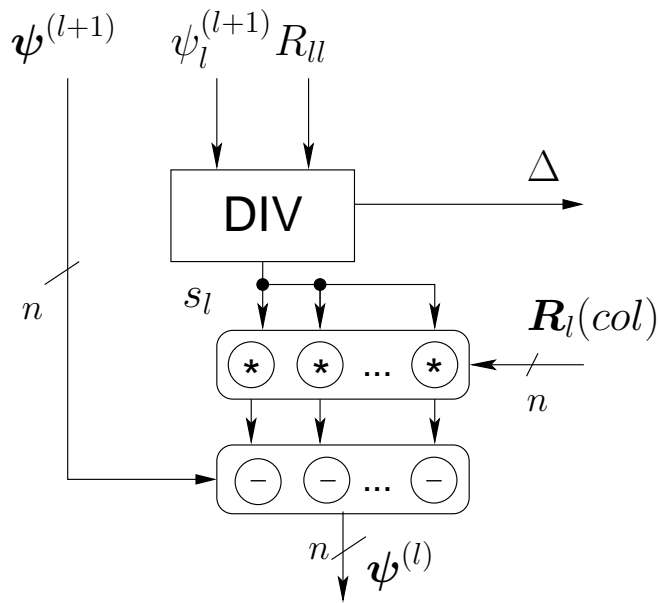}
    \caption{U\_psi Unit datapath}
    \label{U_psi}
    \end{center}
\end{figure}

In order to find the value of $s_l$ able to
minimize $|\psi_l^{(l+1)}-R_{ll}s_l|$ , \textbf{U\_psi} unit (shown in Figure~\ref{U_psi}) 
receives as inputs the
$\psi_l^{(l+1)}$ derived at the upper tree level, together with the
$l$-th diagonal element of matrix $\boldsymbol{R}$. The result of the division 
$\psi_l^{(l+1)} / R_{ll}$ is approximated to the closest odd integer.
This approximation is equivalent to the selection of the closest
point in a $Q$-PAM  constellation. 

The resulting value
directly provides the desired $s_l$ for the analyzed node.
The new $\psi_i^{(l)}$ values are then evaluated in parallel, to be used at the 
lower tree level. 

%% In the notation of Figure~\ref{U_psi},
%% the $\boldsymbol{\psi}^{(l)}$ vector contains the $n$ $\psi_i^{(l)}$
%% values and the update operation can be expressed as
%% \begin{equation}
%% \boldsymbol{\psi}^{(l)}= \boldsymbol{\psi}^{(l+1)}- \boldsymbol{R}_l s_l
%% \label{boldpsieq}
%% \end{equation}
%% \noindent The initial value is given by 
%% $\boldsymbol{\psi}^{(n+1)}=\boldsymbol{\tilde{y}}$.
Vector $\boldsymbol{\psi}^{(l)}$ is stored in a dedicated memory, 
which will be later referred to as \textbf{Psi memory} in
the global architecture given in Figure~\ref{Architecture}. 

The $\Delta$ output in Figure~\ref{U_psi} is defined as
\begin{equation*}
\Delta = s_l - \frac{\psi_l^{(l+1)}}{R_{ll}}
\end{equation*}
\noindent and it represents the correction term to be applied to the division result in order 
to take the closest point in the equivalent PAM constellation. The use of $\Delta$ will be described 
later in this Section.\\

\begin{figure}[ht]
    \begin{center}
    \includegraphics[height=8.0cm]{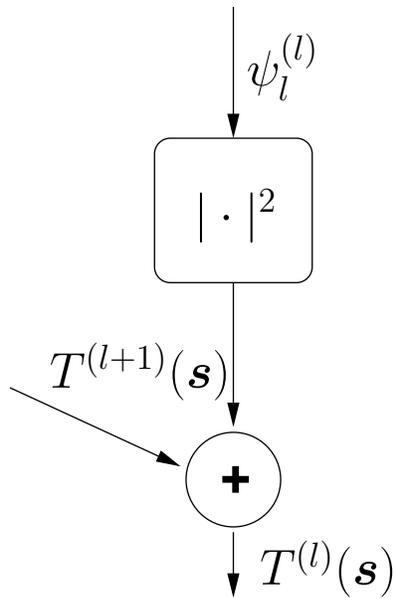}
    \caption{Architecture of the \textbf{Metric\_compute} unit}
    \label{Metric_block}
    \end{center}
\end{figure}

The \textbf{Metric\_compute} unit realizes the second processing step, 
evaluating the new metric $T^{(l)}$ for the selected son.
Figure~\ref{Metric_block} shows the block architecture: from the upper tree level,
$T^{(l+1)}$ is received as input, together with the $\psi_l^{(l)}$ value generated
by \textbf{U\_psi} unit; the obtained $T^{(l)}$ is propagated to the lower tree level.

The described approach, and particularly the use of a division to obtain the 
optimal $s_l$, allows avoiding multiple metric computations; thus it
offers low complexity and, at the same time, flexibility in terms of
supported modulation schemes.
As a matter of fact, a parallel architecture tailored on a given search tree
is able to achieve high processing speed, while
the sequential computation of a single metric at each cycle
makes it easier for the decoder to adapt to different
structures of the search tree, so providing support to multiple modulation
schemes. Similarly to what is done in a software implementation, sequential 
operations compute a single metric at every cycle, so that the same processing 
platform can easily adapt to different structures of the search tree by
simply varying the number of search steps in the tree.

On the other hand, differently from what was implemented in previous
detectors, multiplications cannot be reduced to add and shift
procedures since operands are not fixed and as a consequence general
purpose multipliers have been allocated.

\begin{figure}[t!]
    \begin{center}
    \includegraphics[height=5cm ]{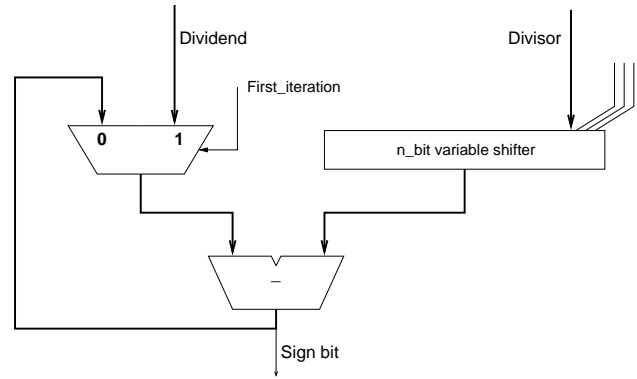} %height=2 cm,
    \caption{Block diagram of the divisor}
    \label{Divisor}
    \end{center}
\end{figure}

It is worth noting that, although the described technique introduces
the division $\psi_l^{(l+1)} / R_{ll}$, only a few values of this ratio
are of interest for the algorithm, those that correspond to the equivalent PAM
constellation points $\pm 1, \pm 3, \ldots$. As a consequence, a
general purpose hardware divisor is not necessary and the required
operation can be executed by means of a simplified component able only to
find the closest integer solution of this division and to determine
if the approximation is by defect or by excess: the first $\log_2 Q$
steps of a successive subtraction divider~\cite{ArithBook} can be
employed to this purpose, where $Q^2$ is the number of signals in
the QAM constellation. This divider has a very simple architecture
that employs only shifts and subtractions; although it tends to be
very slow for a complete division, this solution can be effectively
used when only a few shift and add steps are required. The divider employs a
dichotomous process to find the requested value after $\log_2 Q$
steps. In the block diagram of Figure \ref{Divisor}, the multiplexer
selects the dividend at the first step and the subtraction result 
in the following ones; the $n$-bit variable shifter is
used to shift the divider by a number of positions that changes from
the initial value of $\log_2 Q - 1$ down to 0. The subtractor
returns the result one bit per iteration, starting from
the most significant one.

\subsection{Parallelism and pipelining}

The desired functional flexibility cannot be achieved at the
expenses of processing throughput, but the final architecture must
properly conjugate both features of flexibility and high data rates.
Among effective techniques that can be used to increase throughput,
parallelism and pipelining have been considered. In previous
works, high throughput is obtained resorting to
parallel architectures and two different kinds of parallelism are
usually  employed:
\begin{itemize}
\item Parallelism at the level of tree exploration
\item Parallelism at the level of the metric computation
      for all sons of a given node and in the
      selection of the most probable son.
\end{itemize}
 The first technique can be used only with some suboptimal algorithms \cite{Kbest}
and it becomes unfeasible when optimal algorithms are adopted,
since it requires large amounts of hardware resources.
The second approach
is feasible only with low order QAM modulation schemes as it implies many
concurrent multiplications. Thus these techniques %, for such reasons,
are not viable for the implementation of parametric architectures.
As a consequence, in this work,
the pipelining technique has been investigated.

\begin{figure}[t!]
    \begin{center}
    \includegraphics[width=8cm]{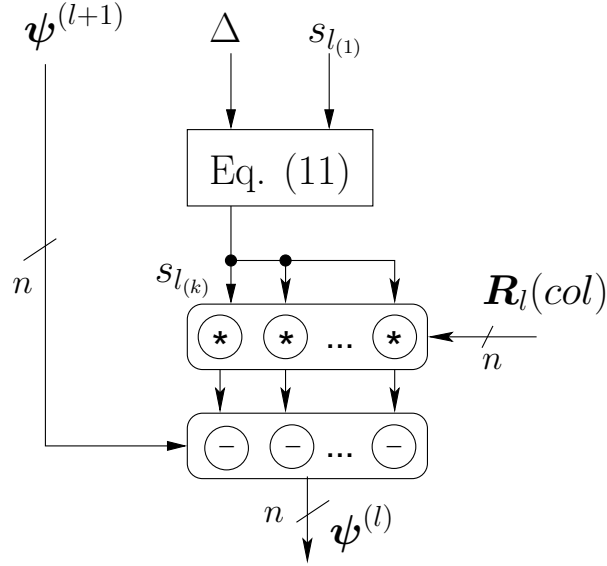} %height=2 cm,
    \caption{Architecture of the U\_psi\_step unit}
    \label{U_psi_step_block}
    \end{center}
\end{figure}

In order to ensure that a new node is expanded at each clock cycle,
a new, alternative metric must be
available also after a pruning operation has taken place. As a
consequence, when the metrics of a given father node are evaluated, two
``candidate'' nodes are concurrently computed: the first one is a direct son
of the current node and it is processed by the \textbf{U\_psi} unit,
while the alternative node, placed at a higher level in the tree, 
is concurrently computed by the 
\textbf{U\_psi\_step} sub-circuit (see Figure \ref{U_psi_step_block}). Both of 
them generate novel $\boldsymbol{\psi}^{(l)}$ values for the next step in the tree traversal.
%%%%%%%%%%%%%%%% rimanegg

\begin{figure}[t!]
    \begin{center}
    \includegraphics[width=0.9\linewidth]{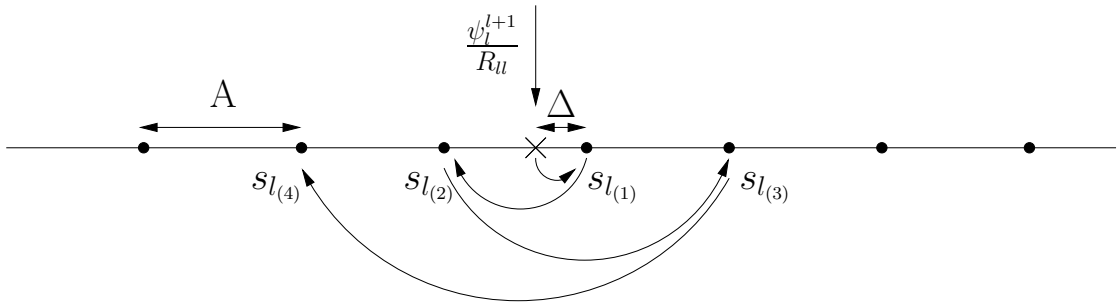} %height=2 cm,
    \caption{Method used to select alternative nodes in U\_psi\_step unit}
    \label{method}
    \end{center}
\end{figure}

\textbf{U\_psi} and \textbf{U\_psi\_step} units share a very similar architecture,
however the latter does not need to perform the division, as the second best choice for $s_l$
(and thus for the alternative node)
can be easily derived as follows. When \textbf{U\_psi} unit computes the division, the result is 
approximated either by defect or by excess to the nearest PAM constellation point:
the best choice for $s_l$ is given by (see Figure~\ref{method})
\begin{equation}
s_{l_{(1)}} = \frac{\psi_l^{(l+1)}}{R_{ll}} + \Delta
\label{delta_def}
\end{equation}
\noindent where $\Delta$ is the correction term provided as output by the 
\textbf{U\_psi} unit (Figure~\ref{U_psi}).

The sign of $\Delta$ is used by \textbf{U\_psi\_step} unit to take the second
(and following) nearest point in the PAM constellation, according to the 
following rule, implemented in the top block of Figure~\ref{U_psi_step_block}
\begin{eqnarray}
s_{l_{(k)}} = s_{l_{(k-1)}} - (-1)^k {\rm sign}(\Delta) \; (k-1) \; A
\label{alternative_node}
\end{eqnarray}
\noindent where $A$ is the distance between two consecutive points and the initial value,
$s_{l_{(1)}}$, is the closest point given in equation~(\ref{delta_def}).

Figure \ref{method} shows the sequence of alternative 
nodes selected at a given tree level, after the occurrence of pruning.
%%%%%%%%%%%%%%%% end rimanegg
Depending on the values assumed by the father node metric, the algorithm
descends along the tree, reaching the son node, or it moves to the 
alternative node on the same level.
It is worth noting that the computations of the $\boldsymbol \psi^{(l)}$ values for both
son and alternative nodes
are performed concurrently with the elaboration of the $T_l$ metric for the
father node.
In other words, while the current metric is computed for the father node,
the next node to be visited is identified choosing
between the son and the alternative node. Additionally, the
related $\boldsymbol \psi_l^{(l)}$ value is computed to be used at the following step
in order to obtain the proper metric $T_{l-1}$ .

This approach also provides a significant speed-up to the inherently
serial SE Sphere Decoding algorithm and  has a limited impact on
complexity.

\subsection{Global architecture}

\begin{figure}[ht]
    \begin{center}
    \includegraphics[width=0.8\linewidth]{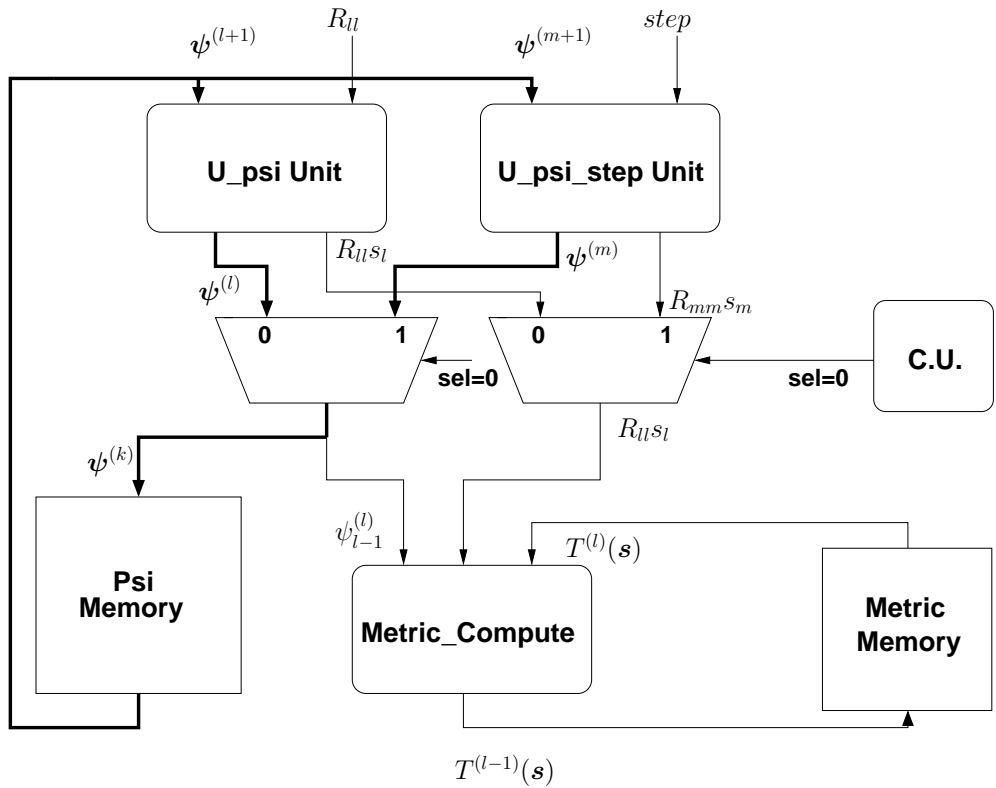}
    \caption{{\it Sphere decoder} block scheme (case of a node expanded in the depth-first mode, with
no pruning). }
    \label{Architecture}
    \end{center}
\end{figure}

The block scheme of the SE tree-traversal circuit showing the architecture 
derived from the design criteria outlined in previous paragraphs is depicted 
in Figure \ref{Architecture}.
Four fundamental processing blocks can be identified in this architecture:

\begin{itemize}
\item \textbf{U\_psi} unit, which selects the most probable son of the current 
node and computes updated $\boldsymbol{\psi}^{(l)}$ through expression~(\ref{boldpsieq}) 
(see also Figure~\ref{U_psi});
\item \textbf{U\_psi\_step} unit, which selects the alternative node to be 
expanded and computes %updated $\psi_l = {\tilde{y_l} - \sum_{j=l+1}^{M_T} R_{lj} s_{j} }$;
for this node the same amount;

\item \textbf{Metric\_compute} unit, which computes metric of the current node 
$T^{(l)}= T^{(l+1)}(\boldsymbol {s}^{(l+1)})+ | \psi_l^{(l+1)}- R_{ll}s_{l}|^2$, 
as in equation (\ref{MetrEq});

\item \textbf{C.U.}, control unit devoted to the proper selection of the
tree search direction.
\end{itemize}

The {\bf C.U.} constitutes the core of the tree traversal algorithm
and it must also carry-out two further tasks: to verify the pruning
condition and, on the basis of this verification, to properly
dispatch data between the other units. Symbols given in Figure~\ref{Architecture}
are related to the case of a node expanded in the depth-first mode, with
no pruning: as a consequence, inputs of the \textbf{Metric\_compute} unit 
are fed with outputs provided by \textbf{U\_psi} block. When a pruning
occurs, multiplexers are switched and metrics related to the alternative node
are selected.

Finally, \textbf{Psi Memory} stores $\boldsymbol{\psi}^{(l)}$ vectors 
from one step to the following one.

%
%%%%%%%%%%%%%%%%%%%%%%%%%%%%%%%%%%%%%%%%%%%%%%%%%%%%%%%%%%%%%%%%%%%%%%%%%%%
\section{Second Hardware Implementation: Flexible Modulation Solution}

\label{HW2}

The capability of managing more than one modulation
scheme in order to adaptively select the most efficient one according to user
needs and channel conditions, is one of the most important requirements of
modern wireless communications systems.
The Golden code, thanks to the non-vanishing determinant property, is
very well suited for such application since it achieves the best performance
independently of the QAM size.
In order to take full advantage of this Golden code feature, an enhanced 
implementation has been realized to allow run-time choice of the modulation 
scheme.

This implementation relies on the same architecture described in the previous 
section, with an additional parameter that allows the run-time selection of 
the constellation.
The requirement of supporting multiple modulation schemes basically impacts on 
the control logic, while the other architecture components remain the same as 
in the first hardware implementation.

At each level of the tree, the C.U., besides the pruning condition
verification, also carries out a second verification task, related to the {\it mapping constraint}:
it verifies if a certain value of $s_l$ still belongs to the specified constellation
and uses this information to drive the processing.

This {\it mapping constraint} must also be taken into account in the
division $\psi_l^{(l+1)} / R_{ll}$. As the number of acceptable values for
this operation depends on the adopted modulation, 
the constellation parameter is used to dynamically drive the
iterations of the dichotomic division algorithm.\\

Although the architecture deals with the implementation of the Golden Code where $n=8$,
it is also scalable in terms of $n$. Increasing the number of transmitting and 
receiving antennas: a larger value of the $n$ parameter
can be set in the VHDL code to synthesize detectors for larger STBcodes. 
Of course a larger $n$ implies a more expensive architecture: particularly
the value of $n$ mainly affects:
\begin{itemize}
\item the number of $\psi$ values to be evaluated in parallel in Figures 6 and 9
\item the depth of the tree 
\item the size of $\psi$ memory.
\end{itemize}
The complexity of processing blocks in Figures~\ref{U_psi} and \ref{U_psi_step_block} 
grows almost linearly 
with $n$; the memory size increases as $n^2$, because $n$ values of $\psi^{(l)}$ have to
be stored for $n$ tree levels. Finally the throughput is expected to decrease
with $n$, since the number of visited nodes grows, but this effect is strongly
dependent also on the supported code.

\section{Synthesis results}
\label{sec:Result}

The first proposed architecture, tailored to process the 16-QAM case,
has been synthesized on both $0.13 \mu$m and $0.25 \mu$m
CMOS Standard Cell technologies, using the Synopsys Version Z-2007.03-SP1; 
synthesis on $0.13 \mu$m technology has been performed
for the second flexible architecture.
A commercial low-power library has been chosen.

In order to enable the direct comparison with existing hardware realizations~\cite{ethvlsi}, I and II ASIC,
\cite{lundvlsi}, a 16 bit datapath has been chosen and the overall decoder has
also been simulated with the uncoded 4$\times$4 MIMO system and throughput
figures reported in Table~\ref{SynthTable} refer to this configuration.
\begin{table*}[t]
\renewcommand{\arraystretch}{1.3}
\centering
\caption{Synthesis results and comparisons (16 bits)}
\vspace {1pt}
\label{SynthTable}
\begin{tabular}{|c||c|c|c|c|c|c|} \hline
& & & & \multicolumn{3}{c|}{This work}\\\hline
Reference & ASIC-I \cite{ethvlsi} & ASIC-II \cite{ethvlsi} & \cite{lundvlsi} & \multicolumn{2}{c|}{PARAMETRIZABLE IMP.} & FLEXIBLE IMPL. \\ \hline\hline

Antennas & \multicolumn{3}{c|} {4$\times$4} &
\multicolumn{3}{c|}{2$\times$2 per two channel uses}\\ \hline
Modulation & 16-QAM & 16-QAM & 16-QAM & \multicolumn{2}{c|}{16-QAM} & 4,16,64-QAM\\ \hline
Detector & depth-first & & K-best & \multicolumn{3}{c|}{depth-first}\\
& sphere & sphere & sphere & \multicolumn{3}{c|}{sphere} \\ \hline
BER Perf. & ML & Close to ML & Close to ML & \multicolumn{3}{c|}{ML} \\ \hline
Tech. [$\mu$m]& 0.25 & 0.25 & 0.35 & 0.25 & 0.13 & 0.13 \\ \hline
Core Area [GE] & 117K & 50K & 91K & 56K & 45K & 55K \\
 & +preproc. & +preproc. & +preproc. & +preproc. & +preproc. & +preproc.\\ \hline
Max. Clock & 51 MHz & 71 MHz & 100 MHz & 109 MHz & 250 MHz & 217 MHz \\ \hline
Throughput & 73 Mbps & 169Mbps & 52 Mbps & 73 Mbps & 167 Mbps & 146 Mbps\\
& @SNR=20 dB & @SNR=20 dB & & @SNR=20 dB & @SNR=20 dB & @SNR=20 dB \\ \hline
\end{tabular}
\end{table*}

The comparison of the described architectures to existing implementations tend 
to be quite difficult to carry out, because different approaches have been
adopted: particularly, our solution implements the ML detection algorithm by means of 
a serial architecture, while the first ASIC in \cite{ethvlsi} maps the same algorithm
onto a parallel structure and the second ASIC in \cite{ethvlsi} makes use of a
serial scheme to realize a close to ML algorithm. These differences must
be carefully evaluated while reading results in Table~\ref{SynthTable}.\\

Comparing the parameterizable architecture to parallel implementations
in Table~\ref{SynthTable}, the solution described in~\cite{lundvlsi} and 
the first ASIC presented in \cite{ethvlsi},
it can be observed that a single metric computation is performed at
each cycle, instead of multiple parallel metric computations. This
characteristic justifies both the reduced complexity and the inherent
flexibility of the proposed architecture.
At the same time, thanks to the adopted pipelined architecture, a
remarkable average decoding throughput is achieved without any 
highly specialized structure.

Implementation cost is slightly higher than for the second ASIC 
proposed in~\cite{ethvlsi}, where a serial approach is also adopted, in conjunction with 
a close to ML algorithmic approach. 

On the other hand, the flexible implementation in the last column of Table~\ref{SynthTable} 
prove the limited complexity and performance overhead associated to the 
capability of dynamically adapting to different modulations (4-,
16- and 64-QAM).\\

Finally, the results presented in Section~\ref{FIX} on the finite precision
analysis of the decoding algorithm have been exploited to derive
additional post synthesis figures for the flexible architecture:
these results, referred to different datapath widths, are given in
Table~\ref{Table:DP Width Synth}. A total of 14 bits are enough for
the 64-QAM modulation (6 bits for the integer part and 8 for the
fractional one) and the two saved bits grant a complexity reduction
of 8 Kgates.

\begin{table}
\centering
\caption{Different datapath width synthesis results}
\begin{tabular}{|c||c|c|c|c|} \hline
DP Width&Area[kG]&Period[ns]&Freq.[MHz]&Through.[Mbps]\\ \hline\hline
12 & 41 & 4.3 & 232 & 155 (16-QAM)\\ \hline
14 & 47 & 4.45 & 224 & 150 (16-QAM)\\ \hline
16 & 55 & 4.6 & 217 & 146 (16-QAM)\\ \hline\end{tabular}
\label{Table:DP Width Synth}
\end{table}

\section{Conclusions}
A novel approach has been presented for the hardware implementation 
of a Sphere Decoder detector: the proposed solution uses a single metric computation per cycle and
is well suited for pipelining, breaking the sequential nature of SD
algorithm.

The main element of novelty of the described approach is in its 
inherent flexibility that makes it suitable for the implementation of
an adaptive modulation scheme.
Two different hardware architectures have been designed: the first 
implementation is a parametrizable one,
while the second is able to adapt on the fly to different modulation
schemes.

The data representation format adopted in both implementations is
based on exhaustive analysis of finite precision effects
collected for 16 and 64 QAM modulations.

Final synthesis results of the proposed architectures are listed in
Table \ref{SynthTable} and show a significant complexity reduction
(approx. $50\%$ for 16 QAM modulation) with respect to parallel
structures. This is mainly due to the single metric computation per
cycle. A remarkable average decoding throughput can be achieved with
both implementations, thanks to the pipelining technique, even if
the hardware was not tailored on a single modulation scheme as all
previously proposed solutions.

\bibliographystyle{IEEEtran}

%\bibliography{IEEEabrv,bare_jrnl}

%\begin{biography}[{\includegraphics[width=1in,height=1.25in,clip,keepaspectratio]{mshell}}]{Michael Shell}
% where an .eps filename suffix will be assumed under latex, and a .pdf suffix
% will be assumed for pdflatex; or if you just want to reserve a space for
% a photo:

%\begin{biography}{Michael Shell}
%\end{biography}

%\begin{biographynophoto}{John Doe}
%\end{biographynophoto}

% insert where needed to balance the two columns on the last page
%\newpage

%\begin{biographynophoto}{Jane Doe}
%\end{biographynophoto}


%\vfill



\end{document}